\documentclass[a4paper,fleqn]{cas-sc}

\usepackage[authoryear]{natbib}
\usepackage{amsmath,amssymb}
\usepackage{graphicx}
\usepackage{bm}
\usepackage{booktabs}
\usepackage{tikz}

\DeclareRobustCommand{\circlednum}[1]{%
  \tikz[baseline=(char.base)]{%
    \node[shape=circle, draw, inner sep=1pt] (char) {\scriptsize #1};
  }%
}

\begin{document}
\let\WriteBookmarks\relax
\def\floatpagepagefraction{1}
\def\textpagefraction{.001}

\shorttitle{Reconfigurable cyber-physical FSI framework}

\shortauthors{Zhang et~al.}

\title[mode=title]{A reconfigurable multi-axis cyber-physical framework for multi-regime fluid--structure interaction experiments}

\tnotetext[1]{This project is funded by Iowa Energy Center.}

\author[1]{Zihan Zhang}[type=editor,
                        auid=000,bioid=1,
                        orcid=0009-0002-2797-9549]
\cormark[1]
\ead{zihan34@iastate.edu}

\credit{Demonstration - Section 4.1 \& 4.2, Software - Section 4.1.2, Writing - Original draft preparation}

\author[1]{Alex Sorensen} [style=chinese]
\ead{alex.sorensen97@gmail.com}

\credit{Experiment setup - Section 3}

\author[1]{Qimin Feng} [style=chinese, orcid=0009-0002-1035-7313]
\ead{qmfeng11@iastate.edu}

\credit{Demonstration - Section 4.3}

\author[1]{Orion Roberts} [style=chinese, orcid=0009-0007-3745-8188]
\ead{oroberts@iastate.edu}

\credit{System validation - Section 3.2}

\author[1]{Linhao Jin} [style=chinese, orcid=0009-0000-6005-3575]
\ead{linhao@iastate.edu}

\credit{System validation - Section 3.1}

\author[1]{Qiang Zhong} [style=chinese, orcid=0000-0002-8435-5938]
\ead{qzhong1@iastate.edu}

\credit{Conceptualization of this study, Demonstration - Section 4.4, Writing - Review \& Editing}

\affiliation[1]{organization={Department of Mechanical Engineering, Iowa State University},
                city={Ames},
                postcode={50011},
                state={Iowa},
                country={United States}}

\cortext[1]{Corresponding author}

\begin{abstract}
Fluid--structure interaction (FSI) experiments are typically built around mechanical dynamics and constraints imposed by the physical apparatus, so changing mass, stiffness, damping, or allowable motion often requires hardware reconfiguration. Here we present a reconfigurable cyber-physical framework in which these properties are instead assigned through software-defined dynamics. The system provides three translational and one rotational degree of freedom, each independently configurable as prescribed, load-responsive, or locked, with operating roles that can also be reassigned during a running experiment. Measured forces and torques are incorporated into real-time virtual dynamic models, while a common supervisory architecture coordinates multi-axis motion, mode switching, synchronized data acquisition, and diagnostic positioning.
The prescribed-motion pathway is validated using a pitching hydrofoil by comparison with published thrust and power scaling trends, while the load-responsive pathway is evaluated using an active-heave/passive-pitch benchmark that reproduces the expected frequency-dependent resonant response over the tested conditions. The same platform is then reconfigured for intra-cycle active--passive pitching, coordinated vertical-axis turbine-surrogate motion, force-driven passive surge, and automated multilayer stereoscopic particle image velocimetry. These results demonstrate that distinct FSI boundary conditions and measurement requirements can be implemented within a common motion, sensing, and control architecture. By treating mechanical roles and constraints as software-defined experimental variables, the framework provides a reusable basis for reconfigurable FSI experiments without redesigning the underlying platform for each application.

\end{abstract}



\maketitle

\section{Introduction}\label{sec:introduction}

Fluid--structure interaction (FSI) experiments depend sensitively on the mechanical impedance and constraint conditions of the tested structure, including its effective mass, inertia, damping, stiffness, and allowable degrees of freedom. In conventional experiments, these properties are typically adjusted through physical modifications to the apparatus, such as replacing springs, dampers, ballast masses, counterweights, or mechanical linkages between trials~\citep{duarte2019experimental,wang2024experimental,duarte2021experimental}. Such approaches can reproduce a wide range of structural configurations, but systematic parameter studies often require repeated hardware changes and are therefore discrete and time-consuming. Changing a physical component may also alter other characteristics of the system, including alignment, friction, structural mass, and mechanical constraints, making it difficult to vary one dynamic parameter independently.

Cyber-physical systems (CPSs) provide an alternative framework in which selected structural dynamics are implemented through real-time computation rather than solely through physical impedance elements~\citep{lee2011virtual,mackowski2011developing}. In these systems, measured fluid forces and moments are incorporated into a feedback loop that either generates actuator force commands or drives a numerical dynamic model whose response is converted into motion commands. Effective mass, damping, stiffness, and constraint conditions can therefore be modified in software while retaining the same underlying physical apparatus and hydrodynamic environment~\citep{su2019energy,su2019resonant}. This separation between the physical test apparatus and the prescribed structural dynamics provides a means of exploring FSI parameter spaces that would otherwise require repeated mechanical reconfiguration.

Existing cyber-physical FSI experiments generally employ two control strategies. Force-command systems compute virtual restoring or damping forces and apply them directly through an actuator, as demonstrated in studies of vortex-induced vibration and other fluid--structure interaction problems~\citep{lee2011virtual,onoue2016characterization}. Position-command systems instead use measured fluid forces or moments as inputs to a numerical structural model and command the resulting displacement response~\citep{mackowski2011developing,su2019energy}. The latter formulation allows the virtual dynamics to be partially decoupled from the physical inertia and stiffness of the experimental apparatus, at the expense of finite sensing, computation, and actuation delay. These approaches have enabled studies of leading-edge vortex dynamics, elastically mounted foils, energy harvesting, and aeroelastic response~\citep{onoue2016characterization,su2019resonant,mackowski2017effect,fagley2016cyber,zhu2020nonlinear}. In parallel, robotic facilities have demonstrated coordinated prescribed motion over multiple degrees of freedom~\citep{fan2019robotic,moreau2024design}, while generalized CPFD formulations and subsequent facilities have established the feasibility of virtual dynamics involving multiple structural coordinates~\citep{mackowski2011developing,waghela2018control}. Thus, neither multi-axis actuation nor multi-degree-of-freedom virtual dynamics alone constitutes the methodological distinction of the present work.

The remaining opportunity is to integrate these capabilities within a common experimental architecture whose mechanical roles can be reassigned according to the requirements of different FSI problems. Existing implementations typically demonstrate subsets of these functions for particular experimental configurations: prescribed multi-axis motion, force-driven passive response, constrained motion, automated experiment execution, or external flow diagnostics. A broadly reusable platform would instead allow individual degrees of freedom to be assigned independently to prescribed, load-responsive, or fixed behavior; permit these roles to coexist and, when required, change during a running experiment; and coordinate the resulting motion with synchronized load acquisition and flow-field measurements. Such an architecture would shift the experimental configuration itself from a largely hardware-defined quantity toward a software-defined variable.

To realize this framework, we developed a four-axis water-tunnel CPS comprising streamwise ($X$), transverse ($Y$), and vertical ($Z$) translation together with rotation ($\Theta$) about the vertical axis. Each axis can be assigned independently to active, passive, or locked operation, allowing prescribed and force-driven degrees of freedom to coexist within the same experiment. Passive motion is implemented using a position-command formulation in which measured force or torque is supplied to a user-defined dynamic model and integrated in real time to obtain the commanded structural response. A six-axis force/torque sensor provides the fluid-loading input, Simulink Desktop Real-Time executes the virtual dynamics and actuator commands, Stateflow supervises axis operating modes and trial progression, and MATLAB provides parameter assignment, automated experiment sequencing, and data acquisition. The Stateflow architecture additionally permits intra-trial reassignment between active and passive operation, while the gantry positioning system can also be incorporated into external diagnostic procedures such as scanning stereoscopic particle image velocimetry.

Before examining broader reconfiguration, the two principal control pathways are evaluated independently. Prescribed-motion performance is assessed using an actively pitching hydrofoil, for which measured thrust and power are compared with established oscillating-foil scaling trends. Force-driven passive operation is evaluated using an active-heave/passive-pitch hydrofoil benchmark, where the measured frequency-dependent pitch response is compared with previously reported resonance behavior~\citep{williamson2019fluid}. These experiments establish the prescribed-motion and virtual-impedance pathways over the operating conditions examined here, while the associated latency, tracking performance, experimental variability, and hardware limits define the range over which the present implementation can be interpreted.

Following validation, the same platform is reconfigured to represent three distinct FSI systems and one flow-diagnostic application. Intra-cycle hybrid pitching combines prescribed and force-driven rotational dynamics within a single oscillation cycle. A vertical-axis turbine surrogate combines coordinated $X$--$Y$ orbital motion with independently prescribed blade rotation. An active-heave/passive-surge configuration couples prescribed lateral motion to force-driven streamwise response through a virtual-mass model. Finally, the $Z$-axis is used as an automated positioning traverse for multilayer stereoscopic PIV~\citep{zhong2021tunable,zhu2025wavenumber,zhong2021aspect}. The first three cases demonstrate how different FSI boundary conditions can be expressed by reassigning the mechanical roles and coordination of the four axes, whereas the PIV case extends the same motion infrastructure to repeatable flow-diagnostic positioning.

Accordingly, the principal contribution of the present work is not a particular number of controlled axes, but a validated and reusable experimental architecture in which prescribed motion, force-driven response, fixed constraints, intra-trial mode reassignment, automated experiment execution, load measurement, and diagnostic positioning are integrated within a common water-tunnel CPS. The emphasis is therefore on the architecture, validation, operating limits, and transferability of the methodology across different classes of FSI experiments, rather than on complete optimization of the individual demonstration cases.

\section{Experimental setup}\label{sec:setup}
\subsection{Hardware Overview}

\begin{figure*}
\includegraphics[width=\textwidth]{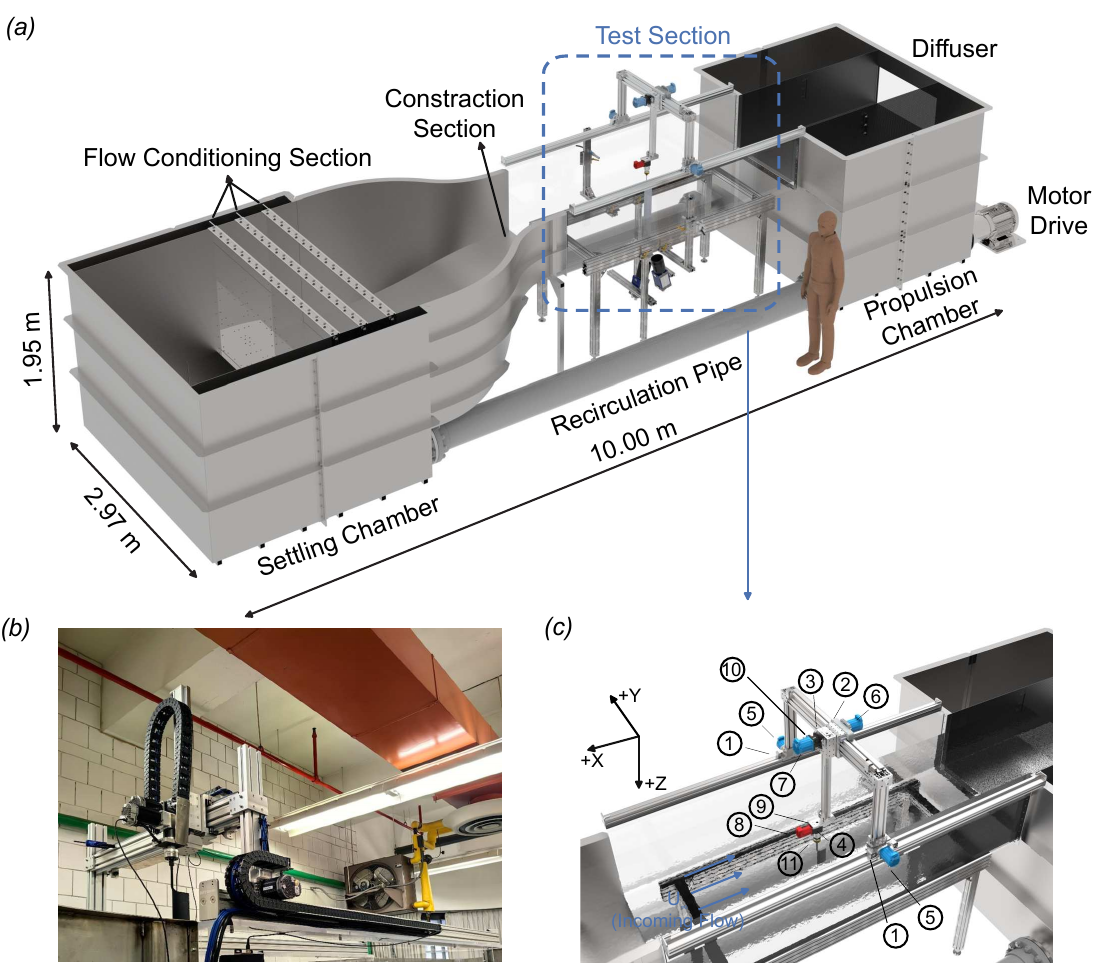}
\caption{Overview of the experimental cyber-physical system (CPS) and recirculating water tunnel. (a) CAD schematic of the complete facility. The CPS is installed above the test section, where experiments are conducted under controlled flow conditions. (b) Photograph of the CPS installed above the water tunnel. (c) CAD schematic of the CPS installed above the test section, with labeled hardware components and coordinate conventions. Actuator (\circlednum{1} $\sim$ \circlednum{4}): \circlednum{1}: $X$-axis; \circlednum{2}: $Y$-axis; \circlednum{3}: $Z$-axis; \circlednum{4}: $\Theta$-axis. Motor (\circlednum{5} $\sim$ \circlednum{8}): \circlednum{5}: $X$-axis; \circlednum{6}: $Y$-axis; \circlednum{7}: $Z$-axis; \circlednum{8}: $\Theta$-axis. \circlednum{9}: Encoder. \circlednum{10}: Z-axis brake. \circlednum{11}: Six-axis Force/torque sensor.}
\label{fig:1}

\end{figure*}

The cyber-physical system (CPS) combines a four-axis gantry, synchronized motion control, force/torque sensing, and real-time data acquisition in a recirculating water tunnel. Figure~\ref{fig:1}(a) shows the complete tunnel, including the motor drive, propulsion chamber, recirculation pipe, settling chamber, flow-conditioning section, contraction, test section, and diffuser. The installed CPS and its labeled hardware components and coordinate definitions are shown in Figs.~\ref{fig:1}(b) and \ref{fig:1}(c), respectively.

Experiments were conducted in a closed-loop water tunnel with a test section measuring approximately $3\,\mathrm{m} \times 1\,\mathrm{m} 
\times 0.5\,\mathrm{m}$ (length $\times$ width $\times$ height). The tunnel was capable of producing uniform freestream velocities up to $1\,\mathrm{m/s}$ within the test section. The gantry assembly was mounted directly above the water tunnel and spanned the width of the test section to provide positioning and motion control of the experimental payload. The structure was constructed using modular aluminum extrusion framing to maintain mechanical rigidity while allowing rapid reconfiguration of experimental hardware.

The CPS provides four independently controlled axes of motion: streamwise translation ($X$-axis), transverse translation ($Y$-axis), vertical translation ($Z$-axis), and rotation about the vertical axis ($\Theta$-axis). The coordinate convention is shown in Fig.~\ref{fig:1}(c). The positive $X$-direction is opposite to the incoming flow, the positive $Z$-direction points vertically downward, and the $Y$-direction completes a right-handed coordinate system perpendicular to the $Z$--$X$ plane.

The streamwise and transverse axes employed ServoBelt LoopTrack linear actuators driven by ClearPath integrated servo motors. The vertical axis utilized a ServoBelt Standard linear actuator to provide controlled positioning along the water depth direction. The rotational axis was actuated through a ClearPath MCPV 2341S-ELS servo motor coupled to a Neugart WPLE-60 planetary gearbox with a 10:1 reduction ratio. This configuration enabled continuous rotational motion while maintaining sufficient torque capacity and angular positioning accuracy for dynamic experiments.

\begin{table*}[t]
\caption{\label{tab:table1}%
Hardware specifications
}
\centering
\begingroup
\footnotesize
\setlength{\tabcolsep}{3pt}
\begin{tabular*}{\textwidth}{@{\extracolsep{\fill}}lcccc@{}}
\toprule
\textrm{Component}&
\textrm{$X$-axis}&
\textrm{$Y$-axis}&
\textrm{$Z$-axis}&
\textrm{$\Theta$-axis}\\
\midrule
Actuator & \shortstack{ServoBelt LoopTrack\\$\times 2$} & ServoBelt LoopTrack & ServoBelt Standard & \shortstack{Neugart WPLE-60\\gearbox (10:1)}\\
\midrule
Motor & \shortstack{ClearPath\\MCPV 3432P-ELS $\times 2$} & \shortstack{ClearPath\\MCPV 3432P-ELS} & \shortstack{ClearPath\\MCPV 3432P-ELS} & \shortstack{ClearPath\\MCPV 2341S-ELS}\\
\midrule
Stroke / range & $2.7\,\mathrm{m}$ & $0.75\,\mathrm{m}$ & $\sim0.5\,\mathrm{m}$  & \shortstack{continuous\\(gearbox output)}\\
\midrule
Max speed & $4\,\mathrm{m/s}$ & $4\,\mathrm{m/s}$ & $4\,\mathrm{m/s}$ & -\\
\midrule
Transmission ratio & $75\,\mathrm{mm/rev}$ & $75\,\mathrm{mm/rev}$ & $75\,\mathrm{mm/rev}$ & -\\
\midrule
Encoder & - & - & - & \shortstack{US Digital E3\\$20,000\,\mathrm{counts/rev}$}\\
\bottomrule
\end{tabular*}
\endgroup
\end{table*}

Hydrodynamic and structural loads were measured using an ATI Industrial Automation Mini40 IP65 six-axis force/torque sensor mounted inline with the experimental payload. The sensor provided simultaneous measurements of three orthogonal force components and three moment components. Analog outputs from the sensor were sampled at 4000~Hz with a $\pm10$~V output range. Prior to each experiment, sensor bias removal and calibration matrix multiplication were performed according to the manufacturer-recommended protocol to recover calibrated force and torque measurements in the sensor coordinate frame.

All sensing and control operations were coordinated through a National Instruments PCIe-6363 data acquisition system. The DAQ hardware supported multi-channel analog sampling rates up to $1\,\mathrm{MHz}$ and provided synchronized acquisition of force, torque, encoder, and auxiliary sensor signals. Counter-based frequency outputs from the DAQ were used to generate motor-control signals for the servo drives, enabling deterministic synchronization between commanded motion and sensor measurements.

To ensure safe operation during power loss or emergency stop conditions, the vertical axis ($Z$-axis) incorporated a Teknic NEMA 34 motor brake. The brake was controlled through transistor-switched outputs from the DAQ system, allowing software-based engagement and release during operation. This configuration prevented unintended vertical motion of the gantry assembly when the servo system was disabled.

The platform reproduces rigid-body translational and rotational degrees of freedom and does not directly emulate distributed deformation such as bending, twisting, or spatially varying compliance. Its accessible virtual-dynamic parameter range is bounded by the actuator travel and speed listed in Table~\ref{tab:table1}, together with the drivetrain, mounting configuration, and structural rigidity of the gantry and test article. Virtual mass, damping, stiffness, and prescribed-motion parameters must therefore remain within this mechanical operating envelope.

\subsection{Control architecture}\label{sec:control}

Two cyber-physical control strategies are commonly used in FSI experiments: force command~\citep{lee2011virtual,onoue2016characterization} and position command~\citep{mackowski2011developing,su2019energy}. Force-command systems compute and apply the restoring force or torque of the target dynamic model, so the achievable virtual response remains coupled to the physical inertia and force capacity of the apparatus.

The present CPS uses the position-command approach. Measured external loads are supplied to a numerical dynamic model, and the integrated response becomes the closed-loop actuator position command. This allows the prescribed virtual mass, damping, and stiffness to differ from the physical properties of the hardware~\citep{mackowski2011developing,su2019resonant,zhang2026inertial}. The sensing, numerical integration, communication, and actuator response introduce a finite delay, which is characterized in Section~\ref{sec:performance}.

\begin{figure*}
\includegraphics[width=\textwidth]{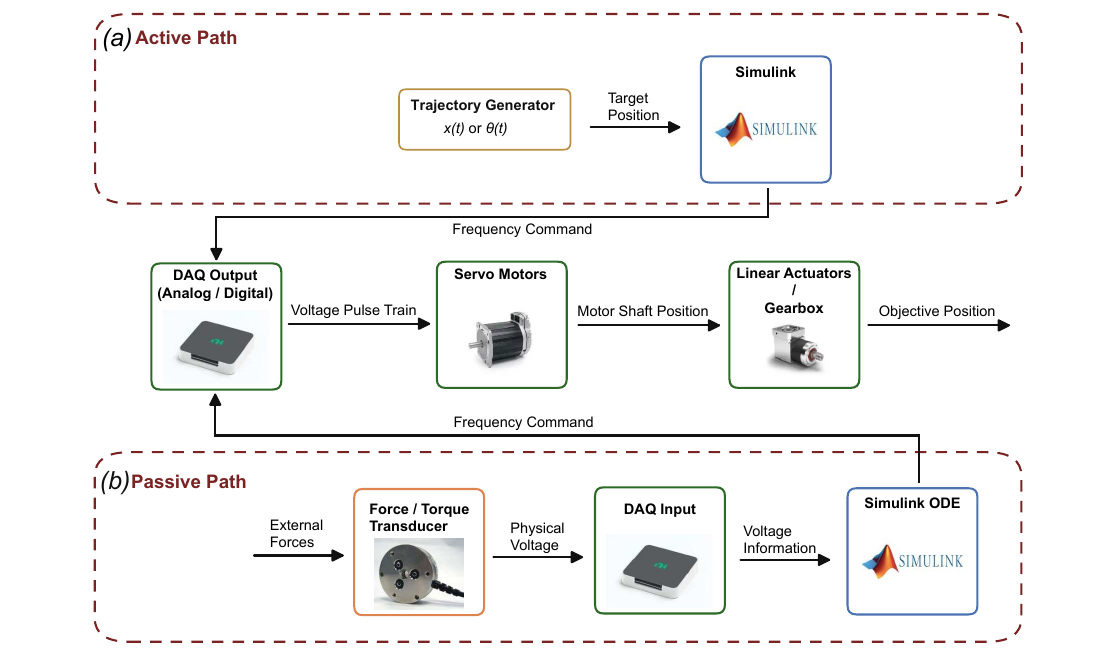}
\caption{Control block diagram showing the signal flow for both (a) Active Path and (b) Passive Path.}
\label{fig:2}
\end{figure*}

Figure~\ref{fig:2} shows the signal paths for active and passive control. In the active path (Fig.~\ref{fig:2}(a)), a user-defined trajectory generator prescribes the desired motion as an explicit function of time, such as $x(t)$ or $\theta(t)$. The target position generated within Simulink is converted into frequency-based motor commands through the DAQ hardware and transmitted to the servo motors. The resulting motor shaft motion is transferred through the corresponding linear actuator or gearbox assembly to produce the commanded physical motion of the experimental axis. The axis thus follows the prescribed kinematics while the system records the fluid and structural loads.

In the passive path (Fig.~\ref{fig:2}(b)), the external force or torque acting on the experimental payload is first measured using the six-axis force/torque transducer. The analog voltage output from the sensor is acquired through the DAQ system and converted into calibrated force and moment signals using the manufacturer-provided calibration matrix. These measured loads are then supplied to a real-time ordinary differential equation (ODE) solver implemented in Simulink. The desired passive dynamics of each degree of freedom are defined through second-order equations of motion. For translational motion, the target behavior is described by:
\begin{equation}
  m\ddot{x} + b\dot{x} + kx 
  = F_{\mathrm{measured}},
  \label{eq:translational eq}
\end{equation}
while rotational motion is governed by:
\begin{equation}
  I\ddot{\theta} + b\dot{\theta} + \kappa \theta 
  = \tau_{\mathrm{measured}},
  \label{eq:rotational eq}
\end{equation}
where $F_{\mathrm{measured}}$ and $\tau_{\mathrm{measured}}$ are the external fluid force and torque measured by the transducer. The parameters $m$, $I$, $b$, $k$, and $\kappa$ represent the prescribed mass, rotational inertia, damping, translational stiffness, and torsional stiffness of the emulated system. The effective translational mass and rotational inertia are defined as:
\begin{equation}
  m = m_\mathrm{physical} + m_\mathrm{virtual},
\end{equation}
and
\begin{equation}
  I = I_\mathrm{physical} + I_\mathrm{virtual},
\end{equation}
where $m_\mathrm{physical}$ and $I_\mathrm{physical}$ correspond to the actual hardware, while $m_\mathrm{virtual}$ and $I_\mathrm{virtual}$ are user-specified virtual quantities introduced through the control framework.

Because the experimental platform contains neither dedicated spring elements nor appreciable mechanical damping, the desired restoring and dissipative effects are imposed numerically. Consequently, the structural stiffness and damping characteristics are represented entirely by virtual parameters, such that $k = k_\mathrm{virtual}$, $\kappa = \kappa_\mathrm{virtual}$, and $b = b_\mathrm{virtual}$. The resulting equivalent structural force acting on a translational degree of freedom is:
\begin{equation}
  F_\mathrm{structure} = -(m_\mathrm{virtual}\ddot{x} + b\dot{x} + kx),
  \label{eq:structural translational}
\end{equation}
and the corresponding structural torque for rotational motion is:
\begin{equation}
  \tau_\mathrm{structure} = -(I_\mathrm{virtual}\ddot{\theta} + b\dot{\theta} + \kappa \theta).
  \label{eq:structural rotational}
\end{equation}
Rearranging Eqs.~\ref{eq:translational eq} and \ref{eq:rotational eq} using the structural loads in Eqs.~\ref{eq:structural translational} and \ref{eq:structural rotational} gives the real-time equations solved by the CPS:
\begin{equation}
  m_\mathrm{physical}\ddot{x} 
  = F_\mathrm{measured} + F_\mathrm{structure},
  \label{eq:translational acceleration}
\end{equation}
and
\begin{equation}
  I_\mathrm{physical}\ddot{\theta} 
  = \tau_\mathrm{measured} + \tau_\mathrm{structure}.
  \label{eq:rotational acceleration}
\end{equation}

The computed acceleration $\ddot{x}$ or $\ddot{\theta}$ is integrated numerically to obtain the corresponding velocity and target position. The target position is converted into a motor-frequency command and transmitted through the actuator chain, producing the physical motion associated with the prescribed virtual dynamics and measured external load.

The software architecture was developed in MATLAB and Simulink using Simulink Desktop Real-Time in ``Run in Kernel'' mode. Separate Simulink subsystems perform sensor calibration, trajectory generation, numerical integration of the virtual dynamics, mode-dependent signal routing, and conversion of target positions into motor-frequency commands. These subsystems execute within the same real-time model so that measured loads, calculated states, commanded positions, and encoder signals share a common time base. Kernel execution provides deterministic timing for synchronized motion control, force acquisition, and numerical integration across all four axes and reduces timing jitter relative to ``Connected I/O'' execution during simultaneous acquisition, closed-loop control, and mode switching~\citep{sorensen2024developing}.

\begin{figure*}
\includegraphics[width=\textwidth]{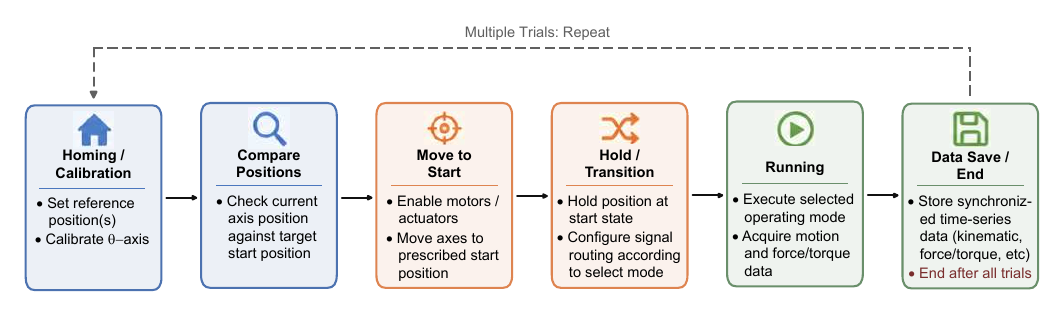}
\caption{Stateflow start-transition diagram for the experimental control sequence.}
\label{fig:3}
\end{figure*}

The experimental workflow is managed through MATLAB scripts, the real-time Simulink model, and a Stateflow supervisory controller. Figure~\ref{fig:3} shows the Stateflow start-transition sequence. Before each experiment, the system enters a homing and calibration state, references the $\Theta$-axis to its predefined zero position, and verifies the rotational calibration. The controller then compares the measured axis positions with the prescribed start conditions. If an offset is detected, the ``Move to Start'' state generates positioning commands for the required axes and holds the remaining axes at their current locations until the payload reaches the specified initial configuration.

Once the target position is reached, the Stateflow controller enters a hold and transition state. The axes remain at the prescribed start condition while the routing logic assigns each axis to active, passive, or locked operation. For a passive axis, the measured displacement, velocity, and acceleration at the transition initialize the corresponding virtual dynamic model. For an active axis, the trajectory generator is initialized at the prescribed phase and start position. The supervisor then enables the selected Simulink subsystems and transfers their outputs to the servo-control path without introducing a displacement discontinuity.

After configuration, the controller enters the running state. Force and torque signals, encoder positions, commanded trajectories, operating-mode flags, and the displacement, velocity, and acceleration states of each virtual model are logged synchronously. The running state also applies the prescribed trial duration and any phase- or condition-dependent switching events. At the end of each trial, the controller transitions to the data-save state, extracts the recorded time series from the Simulink logger, and writes them to file. The sequence then returns to the position-comparison stage for the next condition or terminates after the prescribed parameter sweep.

A supervisory MATLAB script outside the real-time model defines the multi-test sequence. It loads the water-tunnel setting, axis modes, trajectory parameters, virtual mass, damping, and stiffness, together with the trial duration and number of repeats. The script launches and monitors the simulation, checks completion of the Stateflow sequence, extracts the synchronized data, and stores each trial with its parameter set. It then updates the parameters and restarts the control sequence for the next configuration, allowing the same acquisition and file structure to be used across the parameter sweep.

\subsection{Operating modes}\label{sec:modes}

\begin{figure*}
\includegraphics[width=\textwidth]{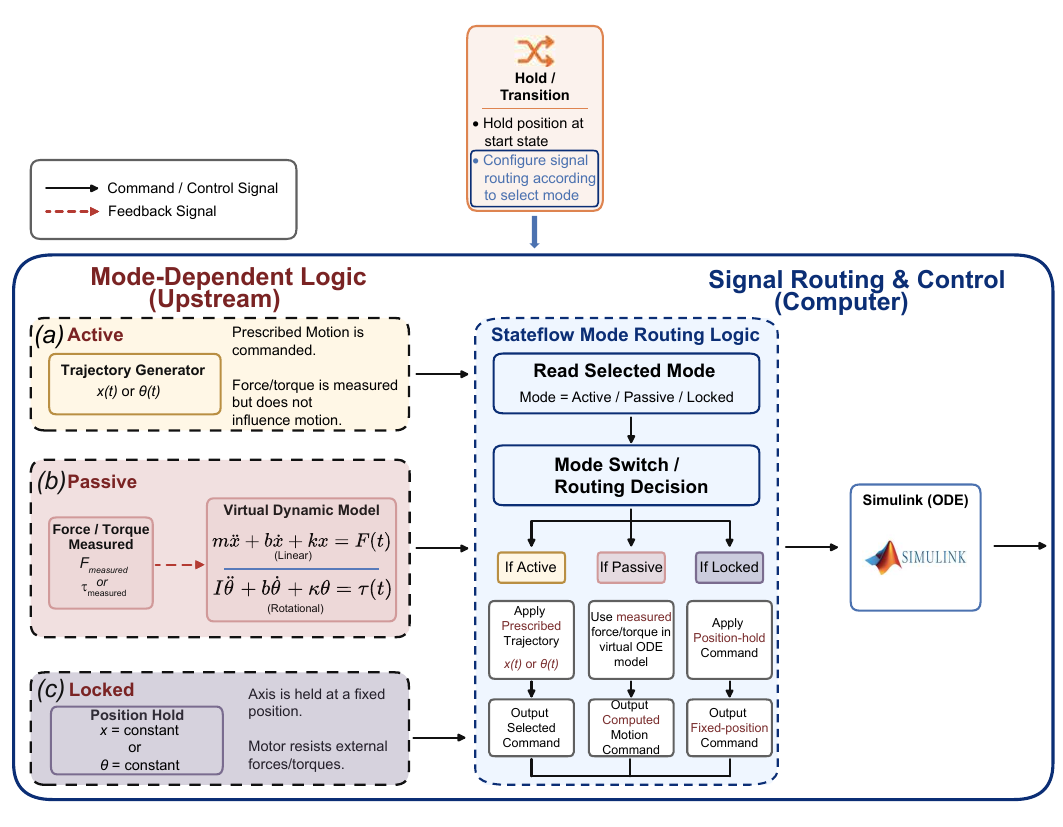}
\caption{Signal routing for (a) active, (b) passive, and (c) locked modes.}
\label{fig:4}
\end{figure*}

Each motion axis is assigned an independently configurable operating mode for dynamically reconfigurable fluid--structure interaction experiments. Figure~\ref{fig:4} shows how mode-dependent motion logic and centralized Stateflow signal routing allow each axis to operate in active, passive, or locked mode. The supervisory controller can also switch an axis between modes during a trial.

The upstream motion logic is separated into three subsystems corresponding to active, passive, and locked operation (Fig.~\ref{fig:4}(a)--(c)). In active mode, the axis follows a trajectory prescribed as a function of time, such as $x(t)$ or $\theta(t)$. The trajectory generator supplies displacement, velocity, and acceleration signals, while the displacement becomes the target sent to the servo-control layer. Force and torque are recorded synchronously but do not affect the commanded motion. This mode is used for forced-motion experiments, prescribed oscillations, coordinated multi-axis trajectories, and traverse operations.

In passive mode, the command results from real-time integration of a virtual dynamic model driven by the measured load. Translational and rotational responses follow Eqs.~\ref{eq:translational acceleration} and \ref{eq:rotational acceleration}, respectively. At each real-time step, the calibrated force or torque is combined with the virtual restoring, damping, and inertial terms, and the resulting acceleration is integrated to update velocity and displacement. The displacement becomes the servo position command, while all calculated states are retained in the data record. The effective mass, damping, and stiffness can therefore be changed in software while the physical hardware remains fixed.

In locked mode, the axis position is maintained at a fixed value using closed-loop position-hold control. The servo system remains enabled and actively resists external disturbances in order to maintain the prescribed position. This operating mode is used when a degree of freedom must remain constrained during an experiment, such as suppressing unwanted structural motion, or isolating the dynamics of another axis.

\begin{table*}[t]
\caption{\label{tab:table2}%
Summary of mode assignments for each demonstration case.
}
\centering
\begin{tabular*}{\textwidth}{@{\extracolsep{\fill}}lcccc@{}}
\toprule
\textrm{Demonstration}&
\textrm{X-axis}&
\textrm{Y-axis}&
\textrm{Z-axis}&
\textrm{$\Theta$-axis}\\
\midrule
EH baseline & Locked & Passive (heave) & Locked & Active (pitch)\\
\midrule
EH hybrid & Locked & Passive (heave) & Locked & Hybrid (pitch)\\
\midrule
VAWT & Active (orbit-x) & Active (orbit-y) & Locked & Active (pitch)\\
\midrule
Aquatic Locomotion & Passive (surge) & Active (heave) & Locked & Locked\\
\midrule
Scanning PIV & Depend on cases & Depend on cases & Active (scan) & Depend on cases\\
\bottomrule
\end{tabular*}
\end{table*}

The Stateflow logic routes one of these three signals to each axis: a prescribed trajectory for active mode, the integrated virtual-dynamic response for passive mode, or a fixed-position command for locked mode. Mode selection is independent across the four axes, so prescribed, load-driven, and constrained degrees of freedom can coexist in one trial. The dynamic models remain separate upstream while sharing the same downstream servo-control path, motor-command conversion, position limits, and data logger.

Mid-trial switching is handled within the Stateflow routing layer by changing the signal path between the mode-dependent subsystems and the downstream position command. Transitions can be triggered at prescribed phases of an oscillation cycle or by other experimental conditions. For an active-to-passive transition, the current active trajectory state initializes the passive dynamic model, so the passive response begins from the instantaneous displacement, velocity, and acceleration of the prescribed motion. For a passive-to-active transition, the active trajectory reconnects from the current passive displacement, and displacement blending reduces abrupt changes in commanded position. Because the controller commands position, it does not independently match velocity and acceleration during this reconnection; differences in these derivative states can produce a finite transient or residual vibration. The hybrid pitching experiment in Section~\ref{sec:hybrid} uses this procedure to alternate between active and passive pitch within each cycle.

Table~\ref{tab:table2} summarizes the mode assignments used in the validation and demonstration cases.

\subsection{Performance characterization}\label{sec:performance}

\begin{figure}
\centering
\includegraphics[width=0.5\columnwidth]{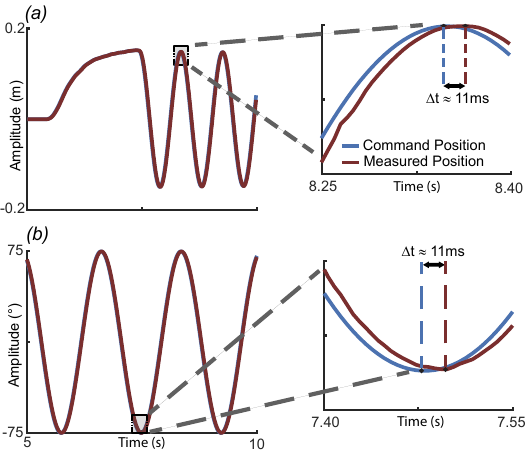}
\caption{Commanded and measured displacement for (a) translational and (b) rotational motion. The enlarged extrema show an approximately 11~ms lag between the measured and commanded positions.}
\label{fig:5}
\end{figure}

We characterized the temporal response and trajectory tracking of the position-command path during translational and rotational motion. Figure~\ref{fig:5}(a) shows a sinusoidal translational trajectory following the initial positioning stage, and Fig.~\ref{fig:5}(b) shows a large-amplitude rotational trajectory. The measured motion follows the waveform of each command, with the largest visible difference occurring as a temporal offset near the extrema. The measured translation and rotation lag their commands by approximately 11~ms.

This end-to-end lag includes signal acquisition, numerical integration, DAQ communication, servo-drive processing, and mechanical actuator response. The real-time model used a response acquisition step size (RAS) of 16~ms; this is the controller execution setting, whereas the 11~ms value is the offset measured from the recorded command and position trajectories. The measured lag is retained when interpreting phase relationships in the experiments below. The tracking difference is concentrated near acceleration reversals, where actuator dynamics and communication latency shift the measured trajectory relative to the command. Repeated trials and cycle-to-cycle variability are reported for the validation cases in Section~\ref{sec:validation}.

Because the four axes differed in mechanical loading and transmission, each used an empirically selected pulse-train command-frequency range that reduced audible vibration and trajectory oscillations while maintaining smooth actuator response. The operating ranges were approximately 1--6~kHz for the streamwise ($X$) axis, 1--3~kHz for the transverse ($Y$) axis, 1--2~kHz for the vertical ($Z$) axis, and 5--10~kHz for the rotational ($\Theta$) axis.

For passive-axis operation, measured force signals occasionally contained high-frequency fluctuations originating from structural vibration and electrical noise. When required, a fourth-order Butterworth low-pass filter with a cutoff frequency of 10~Hz was applied to the force input prior to numerical integration of the passive dynamic equations. The filter operated on signals sampled at 4000~Hz and was used selectively to suppress high-frequency noise while preserving the low-frequency structural dynamics relevant to the experiments.

Additional control stabilization was required for the dual-motor streamwise axis, where synchronization between the two servo actuators prevented torsional loading of the gantry assembly. A proportional--integral--derivative (PID) controller minimized the relative position error between the paired motors during long-travel operation.

Safety and operating-limit logic was implemented within the Simulink real-time model. Motor-disable and emergency-stop commands can be issued directly from the Simulink interface, allowing the user to terminate actuator motion during abnormal operation or experimental setup changes. Position saturation blocks were also applied to each controlled axis using the corresponding mechanical travel limits, preventing commanded trajectories or virtual-dynamic responses from exceeding the allowable range of motion. For the $Z$-axis, the motor brake was kept engaged whenever vertical motion was not required for active or passive operation. This holding condition prevented unintended vertical displacement of the gantry-mounted test article and sensor assembly during locked-axis operation, power interruption, or emergency-stop conditions. Together, the software disable logic, emergency-stop command, axis-specific saturation limits, and $Z$-axis brake provide a layered safety framework for operating the CPS under both prescribed-motion and force-driven passive-response conditions.

\section{System validation}\label{sec:validation}

We tested the two control paths separately. Prescribed pitching was compared with published thrust and power scalings ~\citep{floryan2017scaling,ayancik2019scaling}, and active-heave/passive-pitch motion was compared with an established force-driven FSI benchmark~\citep{williamson2019fluid}.

\subsection{Active-mode validation: pitching-hydrofoil thrust and power}\label{sec:active-validation}

For the active-mode validation experiments, an aluminum NACA 0012 hydrofoil with a chord length of $c = 7.62\,\mathrm{cm}$ and a span of $s = 15.20\,\mathrm{cm}$ was mounted vertically to the six-axis force/torque transducer (Fig.~\ref{fig:1}(a):\circlednum{11}). The hydrofoil was installed with its pitching pivot located at the leading edge ($p/c=0$) and positioned at the centerline of the water tunnel in the transverse ($Y$) direction. Throughout all experiments, the freestream velocity was maintained at $U_{\infty}=0.10\,\mathrm{m/s}$ and monitored using an ultrasonic flowmeter (TUF-2000B). The pitching motion was prescribed through the $\Theta$-axis motor (Fig.~\ref{fig:1}(a):\circlednum{8}), while all remaining axes of the CPS were operated in locked mode.

\begin{figure*}
    \centering  

    \includegraphics[width=\textwidth]{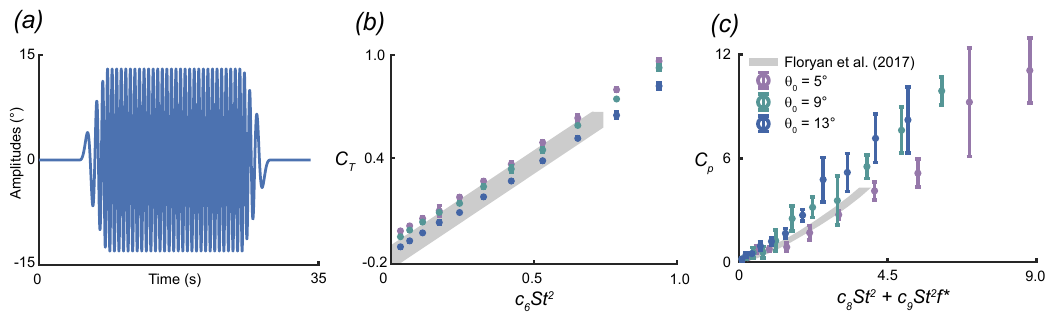}

    \caption{Active-mode validation with prescribed pitching. (a) Pitch trajectory from Eq.~\ref{eq:prescribed pitch} for $\theta_0=13^\circ$ and $St=0.6$. (b) Mean thrust coefficient, $C_T$, as a function of $c_6St^2$. (c) Mean power coefficient, $C_P$, as a function of $c_8St^2+c_9St^2f^*$. Grey regions show the scaling trends of \citet{floryan2017scaling} with $c_6=2.55$, $c_8=7.78$, and $c_9=4.89$; error bars indicate cycle-to-cycle variability.}
    \label{fig:6}
\end{figure*}

The prescribed pitching trajectory was modified from \citet{zhong2024predicting}, which consisted of three ramp-up oscillations, followed by $30$ steady oscillations at the target amplitude and frequency, and finally three ramp-down oscillations to minimize transient loading at the beginning and end of the experiment. The angular motion was prescribed according to:
\begin{equation}
  \theta(t)=
    \begin{cases}
    \dfrac{ft}{6}\theta_0 \sin\left(\pi \dfrac{(ft)^2}{6}\right),
    & 0 \leq t \leq \dfrac{6}{f}, \\[8pt]

    \theta_0 \sin\left(2\pi(ft-3)\right),
    & \dfrac{6}{f} < t \leq \dfrac{36}{f}, \\[8pt]

    \left(\dfrac{ft}{6}-7\right)\theta_0
    \sin\left(\pi \dfrac{(ft-42)^2}{6}\right),
    & \dfrac{36}{f} < t \leq \dfrac{42}{f}.
    \end{cases}
    \label{eq:prescribed pitch}
\end{equation}
where $\theta_0$ is the pitching amplitude and $f$ is the oscillation frequency (Fig.~\ref{fig:6}(a)). The total duration of each experiment was therefore $42/f$. The oscillation frequency was determined from the prescribed Strouhal number:
\begin{equation}
  St = \frac{2fc\sin(\theta_0)}{U_{\infty}}.
\end{equation}

To validate the active-mode performance of the CPS across a broad kinematic range, the Strouhal number was varied from $St = 0.10$ to $0.60$ in increments of $0.05$. For each Strouhal number, the pitching amplitude was prescribed as $\theta_0 = 5^\circ$, $9^\circ$, and $13^\circ$, resulting in a total of 33 unique operating conditions. Three repeated trials were performed for each configuration to evaluate measurement repeatability and trajectory consistency. 

For each operating condition, the system was allowed to evolve through an initial transient period corresponding to the first 5 oscillation cycles in order to reach a statistically periodic state. Following this transient stage, synchronized force, torque, and kinematic measurements were recorded over 20 consecutive oscillation cycles. Phase-averaged quantities were computed using these recorded cycles, and the reported thrust and power coefficients correspond to cycle-averaged values over the statistically periodic regime. Measurement uncertainty was quantified from the cycle-to-cycle variability of the recorded signals and is presented as error bars in the corresponding figures (Fig.~\ref{fig:6}).

The measured hydrodynamic performance was quantified using the mean thrust and power coefficients in order to compare the experimental results with previously reported scaling relationships for oscillating foils \citep{floryan2017scaling,ayancik2019scaling}. The thrust coefficient was defined as:
\begin{equation}
  C_T = \frac{\langle T \rangle}{0.5\,\rho\,U_\infty^2\,c\,s},
\end{equation}
and the power coefficient was defined as:
\begin{equation}
  C_P = \frac{\langle P \rangle}{0.5\,\rho\,U_\infty^3\,c\,s},
\end{equation}
where $\langle \cdot \rangle$ denotes the cycle-averaged value over one oscillation period. Here, $T$ corresponds to the calculated streamwise force acting on the foil:
\begin{equation}
  T(t) = F_x(t)\cos\theta(t) - F_y(t)\sin\theta(t),
\end{equation}
and the instantaneous rotational power was calculated as:
\begin{equation}
  P(t) = \tau_z(t)\dot{\theta}(t),
\end{equation}
where $F_x(t)$, $F_y(t)$, and $\tau_z(t)$ are the instantaneous measured forces/torque and $\dot{\theta}(t)$ is the angular velocity.

Additional correction experiments were conducted to isolate the hydrodynamic loading from non-hydrodynamic contributions. First, stationary drag measurements were performed with the foil aligned with the freestream direction and fully submerged in the water tunnel. The freestream velocity was varied from $U_{\infty}=0.10\,\mathrm{m/s}$ to $0.80\,\mathrm{m/s}$ in increments of $0.10\,\mathrm{m/s}$, resulting in eight drag-test conditions. Each condition consisted of three repeated 20~s measurements, during which the forces acting on the stationary foil were recorded for subsequent drag correction.

Second, inertia-subtraction experiments were performed in air to remove the inertial contribution associated with the oscillating foil and support structure. For these tests, the hydrofoil was operated outside the water while repeating all 33 prescribed kinematic configurations. The measured inertial loads were subsequently subtracted from the corresponding water-tunnel measurements to isolate the hydrodynamic thrust and torque contributions generated by the oscillating foil.

Figure~\ref{fig:6}(b) shows the cycle-averaged thrust coefficient, $C_T$, as a function of $c_6St^2$, and Fig.~\ref{fig:6}(c) shows the cycle-averaged power coefficient, $C_P$, as a function of $c_8St^2+c_9St^2f^{*}$. The grey regions are the scaling trends of \citet{floryan2017scaling}, evaluated with $c_6=2.55$, $c_8=7.78$, and $c_9=4.89$. The present thrust and power measurements follow these trends across the tested Strouhal numbers and pitching amplitudes and are also consistent with the measurements of \citet{ayancik2019scaling}.

The largest deviations from the reference trends occur at high Strouhal numbers and pitching amplitudes, particularly in the power coefficient. Finite-span wake and tip-vortex effects \citep{van2019scaling}, differences in Reynolds number and hydrofoil geometry \citep{floryan2017scaling}, and increased torque uncertainty at large amplitude all contribute to these deviations. Unsteady wake development \citep{buchholz2008wake,floryan2017scaling} and the response delay measured in Fig.~\ref{fig:5} provide additional sources of variation. The error bars in Fig.~\ref{fig:6} report cycle-to-cycle variability among the analyzed cycles. The agreement with the published scaling trends tests the execution of prescribed trajectories and the synchronized acquisition of loads.

\subsection{Passive-mode validation: active heave--passive pitch}\label{sec:passive-validation}
For the passive-mode validation experiments, the aluminum hydrofoil used in the active-mode study was replaced by a 3D-printed ABS NACA 0012 hydrofoil with a chord length of $c=7.62\,\mathrm{cm}$ and a span of $s=22.90\,\mathrm{cm}$. The pitching pivot was at the quarter chord ($p/c=0.25$), matching the experiments of \citet{williamson2019fluid}. Pitch motion was transmitted through a $7\,\mathrm{mm}$ diameter carbon-fiber shaft and a $10\,\mathrm{mm}$ outer-diameter carbon-fiber sleeve that reduced shaft deflection under hydrodynamic loading.

We followed the active-heave/passive-pitch configuration of \citet{williamson2019fluid}. The $Y$-axis prescribed heave, the $\Theta$-axis responded passively through the real-time virtual dynamics, and the remaining axes stayed locked.

The prescribed heaving motion was defined as:
\begin{equation}
  h(t) = h_0\cos(2\pi ft),
\end{equation}
with corresponding velocity and acceleration given by:
\begin{align}
  \dot{h}(t) &= -h_0(2\pi f)\sin(2\pi f t), \label{eq:heave_velocity}\\
  \ddot{h}(t) &= -h_0(2\pi f)^2\cos(2\pi f t). \label{eq:heave_acceleration}
\end{align}
where $h_0$ is the heaving amplitude and $f$ is the heaving frequency. All experiments were conducted at a constant freestream velocity of $U_{\infty}=0.128\,\mathrm{m/s}$, corresponding to a chord-based Reynolds number of $Re = \rho U_{\infty} c / \mu \approx 7.46 \times 10^3$. The heaving amplitude was prescribed as $h_0 = 1.91\,\mathrm{cm}$, corresponding to a non-dimensional heave amplitude of $h_0/c = 0.25$. The oscillation frequency was selected such that the reduced frequency, $f_{\mathrm{red}} = fc/U_{\infty}$, was maintained at $f_{\mathrm{red}} = 0.50$ throughout all experiments.

The passive pitching motion was reproduced through real-time cyber-physical emulation of a virtual torsional spring--mass system. The governing equation for the passive rotational dynamics was:
\begin{equation}
  I\ddot{\theta} = M_p - \kappa \theta + S\ddot{h}\cos\theta,
\end{equation}
where $M_p$ is the hydrodynamic pitching moment measured about the pivot point by the ATI six-axis transducer, $\kappa$ is the virtual torsional stiffness, and $S=m_\mathrm{foil}r$ is the static moment defined by the virtual foil mass and the distance from its center of mass to the pitching axis. The term $S\ddot{h}\cos\theta$ is the inertial moment induced by the prescribed heave acceleration~\citep{williamson2019fluid,duarte2019experimental}. It accounts for the difference between the virtual inertia represented by the CPS and the physical inertia of the printed foil.

The natural frequency of the passive pitching system was determined from:
\begin{equation}
  f_\mathrm{n} = \frac{1}{2\pi}\sqrt{\frac{\kappa}{I}},
\end{equation}
where the rotational inertia $I$ was defined using the user-defined virtual mass. The virtual foil mass was held at $m_\mathrm{foil}=4.70\,\mathrm{kg}$ throughout the passive-mode validation.

To reproduce the passive pitching dynamics reported by \citet{williamson2019fluid}, the frequency ratio $f/f_\mathrm{n}$ was varied from 0.30 to 2.90 in increments of 0.20. Three repeated trials were performed for each configuration. Each trial consisted of an initial 4~s stationary period, followed by 100 heaving cycles and a final 4~s settling period. The passive pitching response was measured using the encoder (Fig.~\ref{fig:1}(a):\circlednum{9}) integrated within the $\Theta$-axis motor assembly.

\begin{figure}
\centering
\includegraphics[width=\columnwidth]{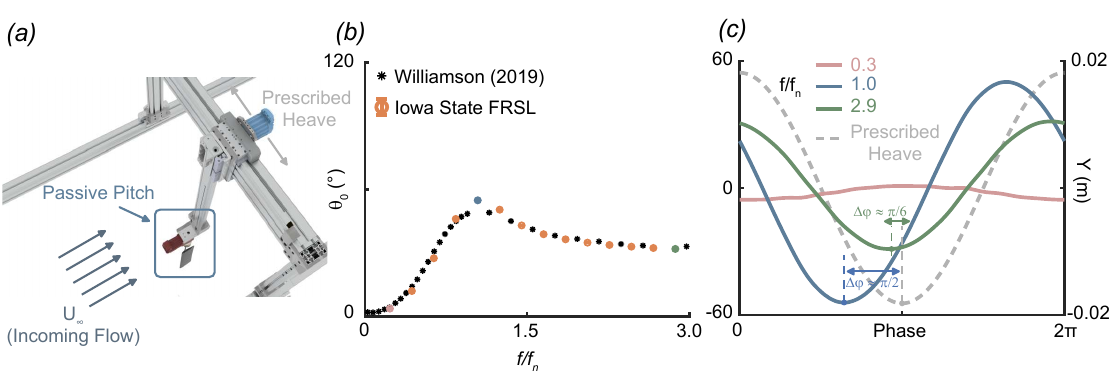}
\caption{Passive-mode validation with active heave and passive pitch. (a) Experimental configuration. (b) Single-sided pitch amplitude, $\theta_0$, as a function of frequency ratio, $f/f_\mathrm{n}$, for $p/c=0.25$, compared with \citet{williamson2019fluid}. Error bars indicate cycle-to-cycle variability. (c) Phase-resolved prescribed heave and passive pitch for $f/f_\mathrm{n}=0.3$, 1.0, and 2.9; annotations indicate the phase offset between the two motions.}
\label{fig:7}
\end{figure}

After the initiation of the prescribed heaving motion, the system was allowed to evolve through an initial transient response before data acquisition began. The first 5 oscillation cycles were excluded from analysis to ensure that the passive pitching dynamics reached a repeatable periodic state. Subsequent passive-response measurements were then acquired over 90 consecutive heaving cycles. The reported passive pitching amplitude, $\theta_0$, was determined from cycle-averaged statistics computed over this periodic measurement window. Experimental uncertainty was evaluated from the cycle-to-cycle variation of the passive response and is presented as error bars in Fig.~\ref{fig:7}.

The passive pitch amplitude increases as the forcing approaches the virtual natural frequency, reaches a maximum near $f/f_\mathrm{n}\approx1$, and decreases at higher frequency ratios (Fig.~\ref{fig:7}(b)). This resonance trend follows the reference measurements of \citet{williamson2019fluid} and the behavior reported for related coupled-foil systems~\citep{su2019resonant,heathcote2007flexible}. The phase-resolved signals in Fig.~\ref{fig:7}(c) also show the accompanying change in phase between prescribed heave and passive pitch. Differences near the peak response are consistent with differences in structural damping, finite-span loading, effective virtual inertia, and measured hydrodynamic moment. The comparison tests the combined virtual inertia, torsional stiffness, inertial coupling, and real-time load-driven motion used in the passive path.

\section{Reconfigurable experimental demonstrations}\label{sec:demonstrations}

Following validation of the prescribed-motion and force-driven response paths, the platform was reconfigured to represent three distinct FSI systems and one flow-diagnostic application. The energy-harvesting configuration combines prescribed pitch with force-driven heave and then introduces phase-dependent switching of the pitching axis between active and passive dynamics. The vertical-axis turbine surrogate maps synchronized $X$--$Y$ translation and independent blade pitch onto an orbital kinematic system, while the locomotion configuration assigns the force-driven response to the streamwise $X$-axis and retains prescribed lateral actuation. In each of these three cases, the same motion, sensing, virtual-dynamic, and supervisory framework is used to realize a different set of FSI boundary conditions by reassigning the roles and coordination of the four axes. The final demonstration extends the platform beyond structural-response emulation by using the $Z$-axis as an automated positioning traverse for scanning stereoscopic PIV. In this case, the CPS does not represent an additional mechanical FSI system; instead, it provides repeatable spatial positioning and synchronization for multilayer flow-field acquisition within the same experimental infrastructure.

\subsection{Flow energy harvesting with hybrid mode switching}\label{sec:hybrid}
\subsubsection{Baseline---active pitch and passive heave}

Semi-passive hydrofoil energy harvesting couples an actively prescribed pitching motion to a heaving response governed by fluid loading and structural impedance~\citep{sorensen2024developing,su2019resonant,zhang2026inertial}. This coupling provides a natural starting configuration for the reconfigurable platform: the $\Theta$-axis supplies the prescribed kinematics, the $Y$-axis evolves from measured force through a virtual spring--mass--damper model, and the remaining axes impose fixed constraints. The resulting active-pitch/passive-heave state also serves as the baseline from which intra-cycle mode switching is introduced in the next subsection.

\begin{figure*}
\centering
\includegraphics[width=0.70\textwidth]{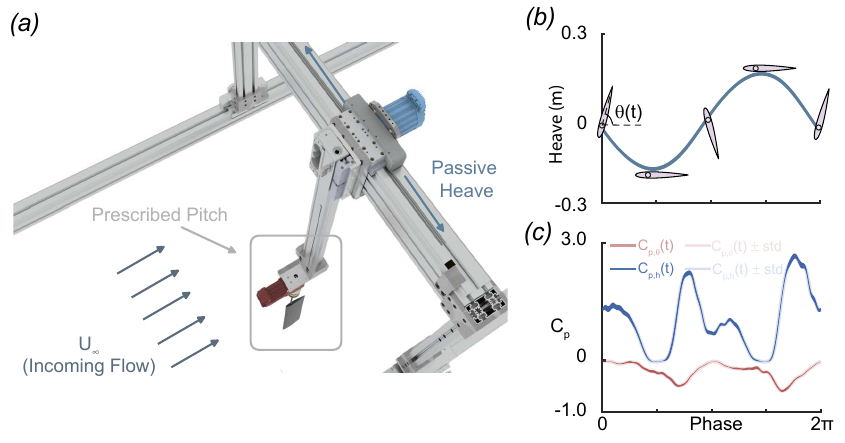}
\caption{Baseline semi-passive energy harvesting with active pitch and passive heave. (a) Experimental configuration and axis assignments. (b) Phase-averaged passive heave displacement, $h(t)$, with foil orientations indicating the prescribed pitch, $\theta(t)$. (c) Phase-averaged pitch and heave power coefficients, $C_{P,\theta}(t)$ and $C_{P,h}(t)$; shaded regions indicate cycle-to-cycle variability.}
\label{fig:8}
\end{figure*}

We configured the CPS for the baseline semi-passive energy-harvesting condition of \citet{su2019resonant}, using the hydrofoil geometry of the present experimental platform. The test model was a NACA 0012 hydrofoil with chord length $c=0.10\,\mathrm{m}$ and span $s=0.38\,\mathrm{m}$. The pitching axis was located at the quarter chord ($p/c=0.25$), and the prescribed pitching amplitude was $\theta_0=75^\circ\approx1.31\,\mathrm{rad}$. The freestream velocity was $U_{\infty}=0.40\,\mathrm{m/s}$, corresponding to $Re=\rho U_{\infty}c/\mu \approx 3.06 \times10^4$. The $\Theta$-axis imposed sinusoidal pitch, the $Y$-axis responded passively through a virtual spring--mass--damper model, and the $X$- and $Z$-axes remained locked. Following \citet{su2019resonant}, the virtual damping ratio and frequency ratio were $b^{*}=1.50$ and $f/f_\mathrm{n}=1$. The virtual mass ratio was $m^*=4.90$, and the prescribed stiffness was calculated by the natural frequency,  $f_\mathrm{n}=\sqrt{k/m}/2\pi \approx 0.64\,\mathrm{Hz}$. The resulting reduced frequency was $f_\mathrm{red}=fc/U_{\infty}=0.144$, within the range identified in previous foil energy-harvesting studies~\citep{zhang2026inertial,deng2015inertial}.

Hydrodynamic forces measured by the six-axis transducer were supplied in real time to the passive heave subsystem, which integrated the prescribed virtual dynamics. The phase-averaged heave and pitch power coefficients, $C_{P,h}(t)$ and $C_{P,\theta}(t)$, are shown in Fig.~\ref{fig:8}(c). Their cycle-to-cycle variation is indicated by the shaded regions.

Throughout the experiment, the prescribed pitching motion remained synchronized with the commanded trajectory, while the heave degree of freedom responded continuously to the measured fluid forcing. The periodic pitch and heave power signals capture the energy exchange associated with the two axis roles. This baseline establishes a repeatable coupled state in which one axis is kinematically prescribed and a second axis is governed by real-time virtual dynamics.

\subsubsection{Intra-cycle active--passive switching}

\begin{figure*}
\centering
\includegraphics[width=0.8\textwidth]{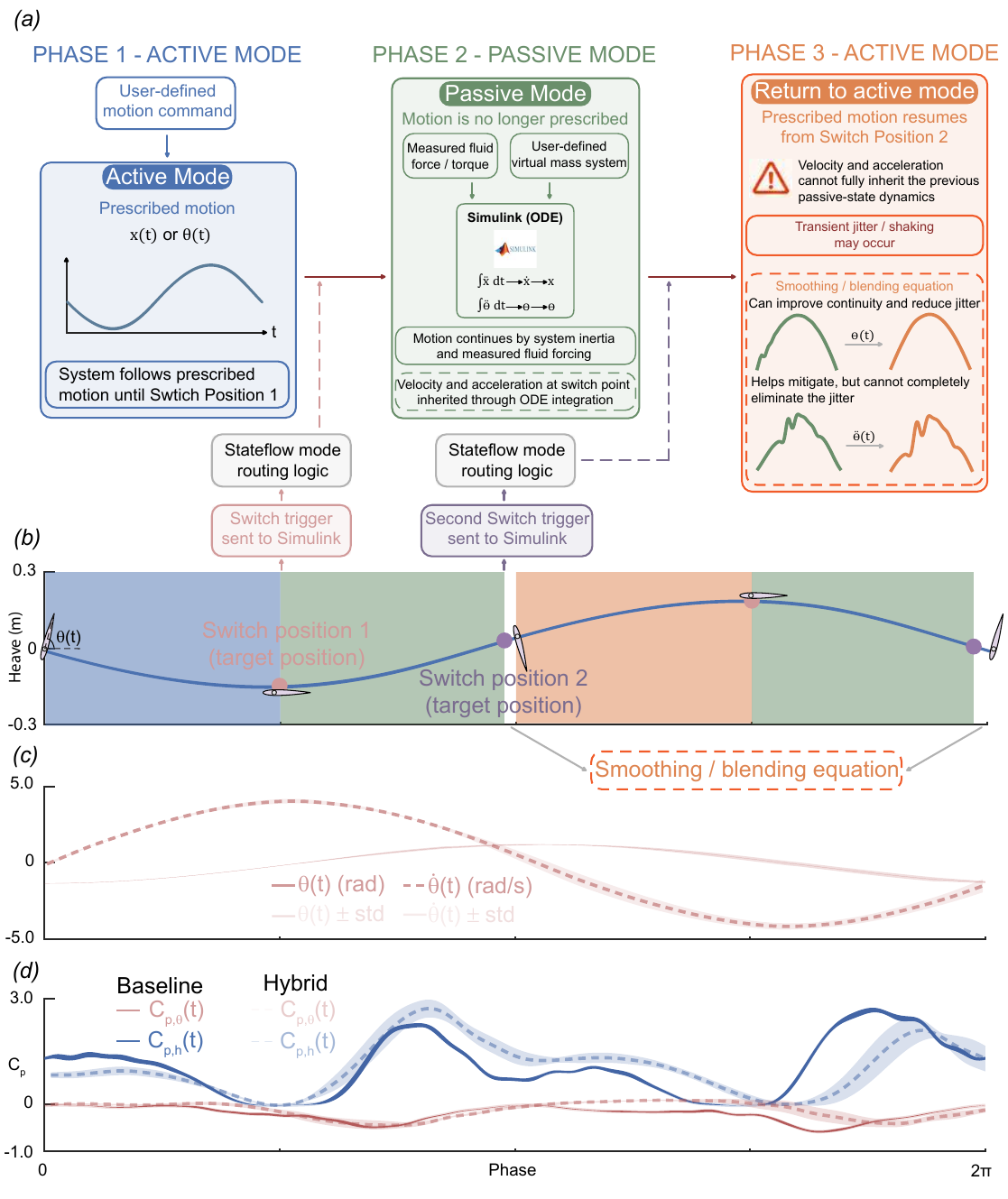}

\caption{
Hybrid active--passive pitching for the semi-passive energy-harvesting configuration.
(a) Stateflow sequence for intra-cycle switching of the $\Theta$-axis from prescribed to passive pitching and back to prescribed motion.
(b) Operating modes and switching locations over one cycle; the $Y$-axis remains passive in heave.
(c) Phase-averaged pitch angle and angular velocity during hybrid operation.
(d) Phase-averaged pitch and heave power coefficients for the baseline and hybrid cases.
Markers denote switching locations, solid and dashed curves denote the baseline and hybrid responses, respectively, and shaded regions indicate cycle-to-cycle variability.
}
\label{fig:9}

\end{figure*}

The baseline configuration assigns a fixed role to each axis for the duration of a trial. Hybrid operation adds a time-dependent mechanical setup: the $\Theta$-axis follows prescribed pitch over selected portions of the cycle and evolves from the measured hydrodynamic moment over the remaining portions. Stateflow routing changes the command source at user-defined phase locations, while the $Y$-axis continues to respond passively in heave. The experiment therefore combines prescribed actuation, two force-driven degrees of freedom, and repeated intra-cycle mode transitions within one periodic state.

The hybrid switching procedure is illustrated schematically in Fig.~\ref{fig:9}(a). During the first active portion of the cycle, the $\Theta$-axis follows the prescribed pitching trajectory used in the baseline active-pitch/passive-heave case. When the first switching condition is reached, the Stateflow controller transfers the $\Theta$-axis from active to passive mode. The passive pitch response is then computed from the measured hydrodynamic moment and the user-defined virtual inertia system. In this active-to-passive transition, the passive ODE is initialized using the instantaneous displacement, velocity, and acceleration from the preceding active trajectory, allowing the passive response to continue from the current motion state. When the second switching condition is reached, the axis is returned to active operation. A displacement-blending procedure reconnects the passive response to the prescribed active trajectory while reducing abrupt changes in the position command.

The blending function used during the passive-to-active transition is a cubic smoothstep function:
\begin{equation}
   w(s)=3s^2-2s^3, \qquad 0\leq s \leq 1,
\end{equation}
where $s$ is the normalized blending coordinate over the prescribed transition interval. The blended command is written as:
\begin{equation}
   \theta_{\mathrm{cmd}}(s)
    =
    w(s)\theta_{\mathrm{act}}(s)
    +
    \left[1-w(s)\right]\theta_{\mathrm{pas}}(s),
\end{equation}
where $\theta_{\mathrm{pas}}$ is the pitch displacement obtained from the preceding passive response and $\theta_{\mathrm{act}}$ is the prescribed active trajectory to which the system returns. This definition gives $\theta_{\mathrm{cmd}}=\theta_{\mathrm{pas}}$ at $s=0$ and $\theta_{\mathrm{cmd}}=\theta_{\mathrm{act}}$ at $s=1$, while the endpoint conditions $w'(0)=w'(1)=0$ reduce abrupt changes in the blended displacement command. Because the CPS is position-command based, the reconnection directly blends displacement rather than independently prescribing velocity and acceleration. The procedure therefore maintains a continuous position command while derivative-state differences appear as finite transition variations.

The baseline semi-passive harvesting configuration was then extended by introducing hybrid operation on the pitching axis. The $Y$-axis remained in passive heave mode throughout the experiment, while the $\Theta$-axis alternated between active and passive operation at prescribed phase locations within each pitching cycle. During the active portions, the pitching motion followed the baseline sinusoidal trajectory. At the first switching phases, $\phi=\pi/2$ and $3\pi/2$, the $\Theta$-axis was transferred from active to passive operation. At the second switching phases, $\phi\approx3\pi/4$ and $7\pi/4$, the axis returned to active operation using the displacement-blending procedure described above. The corresponding switching locations are shown in Fig.~\ref{fig:9}(b).

The measured hybrid response is shown in Fig.~\ref{fig:9}(c). The pitch displacement remains continuous through the switching events, and the pitch velocity remains bounded as the axis reconnects to the prescribed trajectory. The resulting power coefficients are compared with the baseline active-pitch case in Fig.~\ref{fig:9}(d). Reassigning the pitching-axis dynamics over the cycle changes the phase-dependent distribution of pitch and heave power, directly connecting the switching schedule to the balance between prescribed actuation and fluid-driven response. In this configuration, the mechanical setup itself becomes a programmable function of oscillation phase.

\subsection{Vertical-axis wind turbine surrogate}\label{sec:vawt}

The vertical-axis wind turbine (VAWT) surrogate reconfigures the platform from coupled prescribed--passive dynamics to a coordinated three-axis kinematic system. Synchronized $X$- and $Y$-axis translations generate the circular blade orbit, while the $\Theta$-axis independently prescribes blade orientation as a function of orbital phase. This mapping reproduces the coupled translation and pitching that govern the phase-dependent loading and power transfer of a VAWT blade~\citep{strom2017intracycle}, while allowing the orbital path and pitch schedule to be defined independently in software~\citep{li2018optimization}.

\begin{figure*}
\centering
\includegraphics[width=\columnwidth]{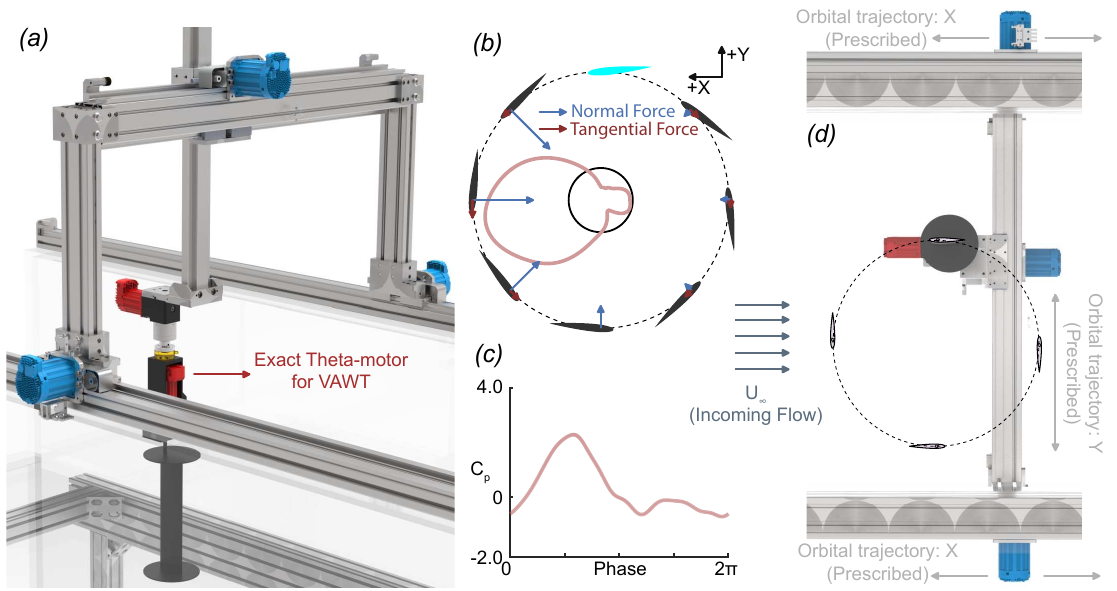}
\caption{Vertical-axis wind turbine (VAWT) surrogate using coordinated multi-axis motion. (a) Side view of the experimental configuration and independently driven $\Theta$-axis. (b) Prescribed $X$--$Y$ orbit, blade pitch, and normal and tangential force directions; the blue foil marks the initial position. (c) Phase-averaged net power coefficient, $C_P(t)$. (d) Bottom view of the configuration; the dashed circle denotes the prescribed orbital path.}
\label{fig:10}
\end{figure*}

Figure~\ref{fig:10} shows the VAWT experiment produced by coordinated multi-axis operation. A single turbine blade was mounted beneath the traverse system, and the $\Theta$-axis motor prescribed blade pitch independently of the orbital translation (Fig.~\ref{fig:10}(a)). The force/torque sensor therefore remained at a fixed orientation relative to its cable while the blade completed the virtual orbit, avoiding cable winding and preserving continuous load acquisition. The blade was a NACA 0018 hydrofoil with chord length $c=0.08\,\mathrm{m}$ and span $s=0.32\,\mathrm{m}$. The prescribed turbine diameter was $D=0.40\,\mathrm{m}$, corresponding to an orbital radius of $R=0.20\,\mathrm{m}$. Experiments were conducted at a freestream velocity of $U_{\infty}=0.20\,\mathrm{m/s}$. The $X$- and $Y$-axes operated in active mode and generated the circular orbital trajectory, the $\Theta$-axis imposed the phase-dependent blade pitch, and the $Z$-axis remained fixed. For the representative case shown in Fig.~\ref{fig:10}, the orbital frequency was $f_{\mathrm{orb}}=0.40\,\mathrm{Hz}$, giving a tip-speed ratio of $\lambda = \omega R/U_{\infty} = 2.51$, where $\omega = 2\pi f_{\mathrm{orb}}$. The three active axes thus define a common orbital phase while retaining independent control of translation and blade orientation.

The prescribed orbital motion and force decomposition are shown schematically in Fig.~\ref{fig:10}(b). The dashed circle represents the blade trajectory generated by the synchronized $X$- and $Y$-axis commands, and the foil orientations indicate the actively prescribed pitch motion at representative orbital phases. The prescribed kinematics can be written in terms of the orbital phase, $\psi(t)=2\pi f_{\mathrm{orb}}t$. The translational commands are:
\begin{equation}
    X(t) = R\sin\psi(t), \qquad
    Y(t) = R\cos\psi(t),
\end{equation}
with $R=0.20\,\mathrm{m}$. For a purely periodic turbine-surrogate rotation, the blade orientation follows the orbital phase, $\Theta(t)=\psi(t)$. When an additional active pitch schedule is imposed, the rotational command becomes:
\begin{equation}
    \Theta(t)=\psi(t)+\theta_0\sin\psi(t),
\end{equation}
where $\theta_0$ is the prescribed pitch amplitude. For the representative case shown in Fig.~\ref{fig:10}(b), $\theta_0=8^\circ$.

The measured hydrodynamic force was decomposed into normal and tangential components relative to the instantaneous orbital path. The tangential component, $F_\mathrm{tan}$, contributes directly to torque production along the orbit, whereas the normal component represents the radial loading on the blade and support structure. A bottom-view schematic of the realized VAWT configuration is shown in Fig.~\ref{fig:10}(d), where the blade follows the circular path relative to the incoming flow, $U_\infty$. The resulting phase-averaged net power coefficient is shown in Fig.~\ref{fig:10}(c). The net power is computed from the balance between the power extracted through orbital motion and the power consumed by blade pitching:
\begin{equation}
  P_\mathrm{net}(t) = F_\mathrm{tan}(t) \cdot V_\mathrm{orbit}(t) - M_{\theta}(t) \cdot \dot{\theta}(t),
\end{equation}
where $V_\mathrm{orbit}$ is the blade orbital velocity, $M_{\theta}$ is the pitching moment, and $\dot{\theta}$ is the pitching angular velocity. The first term quantifies the power transferred from the flow to the orbital motion, whereas the second term accounts for the actuation power required to maintain the prescribed pitch kinematics.

The VAWT case therefore uses the same synchronized axis commands and load acquisition as the preceding FSI configuration, with the measured forces resolved in the moving orbital frame.

\subsection{Active-heave/passive-surge locomotion}\label{sec:locomotion}

Swimming propulsion couples prescribed lateral motion to a streamwise response generated by the resulting unsteady hydrodynamic force~\citep{feng2026passive}. The locomotion configuration assigns these two roles to separate translational axes: the $Y$-axis prescribes lateral heave, and the measured streamwise force drives the $X$-axis through a software-defined virtual mass. The platform thereby converts thrust generated by the oscillating propulsor into a force-driven surge trajectory while retaining direct control of the lateral kinematics.

\begin{figure*}
    \centering  

    \includegraphics[width=\textwidth]{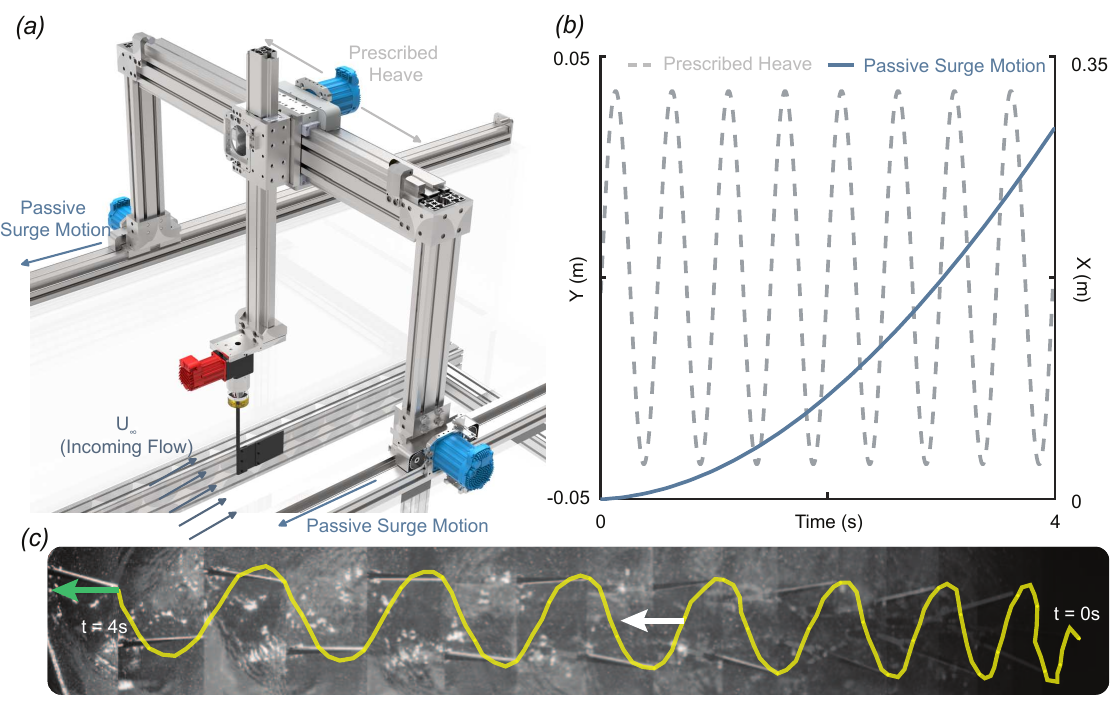}

    \caption{Active-heave/passive-surge locomotion. (a) Experimental configuration with prescribed $Y$-axis heave, force-driven $X$-axis surge, and locked $Z$- and $\Theta$-axes. (b) Prescribed heave and passive surge displacement; dashed and solid curves denote the $Y$- and $X$-axis motions, respectively. (c) Composite images of the resulting two-axis trajectory from $t=0$ to $4~\mathrm{s}$.}
    \label{fig:11}
\end{figure*}

A rigid rectangular plate with a length of $100\,\mathrm{mm}$ and a width of $76.50\,\mathrm{mm}$ was used as a simplified fish-tail model. Figure~\ref{fig:11}(a) shows the active-heave/passive-surge configuration. The model was mounted to the force/torque sensor and tested in the recirculating water tunnel at $U_{\infty}=0.20\,\mathrm{m/s}$. The $Y$-axis prescribed sinusoidal heave with a peak-to-peak amplitude of $0.08\,\mathrm{m}$, corresponding to $\pm 0.04\,\mathrm{m}$ from equilibrium. The forcing frequency was $f=2\,\mathrm{Hz}$, giving a Strouhal number of $St=0.80$. The $X$-axis response was governed by the virtual-inertia model $m_{\mathrm{virtual}}\ddot{x}=F_x$, with $m_{\mathrm{virtual}}=3\,\mathrm{kg}$. For this configuration, the virtual model retained the force--inertia coupling directly; speed-dependent body drag, structural damping, and added-mass variation were set to zero. The $Z$- and $\Theta$-axes remained fixed.

Figure~\ref{fig:11}(b) presents the prescribed heave and force-driven surge over a 4~s interval. The lateral oscillation remains periodic while the measured streamwise force is integrated through the virtual-mass model into the surge response. The composite images in Fig.~\ref{fig:11}(c) trace the resulting two-axis trajectory. Relative to the energy-harvesting configuration, the force-driven axis has moved from transverse heave to streamwise surge, and the virtual dynamics have changed from a spring--mass--damper oscillator to a mass-only model.

\subsection{Stacked multilayer stereoscopic PIV}\label{sec:piv}

Stacked multilayer stereoscopic PIV reconstructs a three-dimensional wake from 2D--3C planes measured as a model moves through fixed laser sheets~\citep{zhong2021tunable,zhu2025wavenumber,zhong2021aspect}. Keeping the cameras and laser sheets fixed preserves a common optical calibration across the scan. Here the $Z$-axis moves the model, and the trial sequence coordinates scan position, prescribed or passive kinematics, and phase-locked acquisition.

\begin{figure*}
\centering
\includegraphics[width=0.98\textwidth]{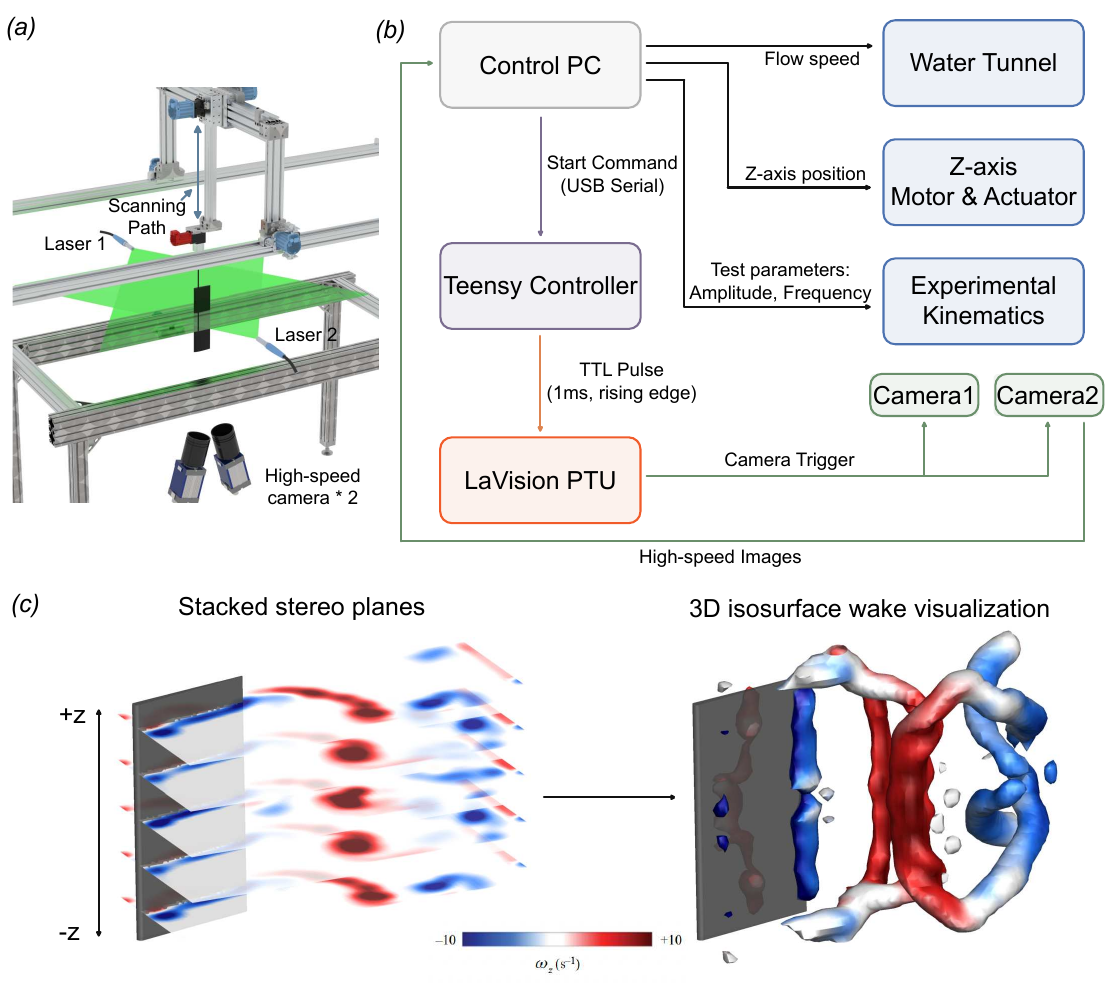}
\caption{Stacked multilayer stereoscopic PIV integrated with the CPS. (a) The $Z$-axis moves the model through fixed laser sheets while two cameras acquire stereoscopic images. (b) Synchronization architecture. The Teensy sends one 1~ms-wide TTL start pulse to the LaVision PTU; the LaVision system controls the subsequent camera timing. (c) Phase-locked stereo-PIV planes and the reconstructed three-dimensional wake.}
\label{fig:12}
\end{figure*}

The scanning-PIV configuration is shown in Fig.~\ref{fig:12}(a). The laser sheets remain fixed in the laboratory frame, while the CPS translates the test model along the $Z$ direction through a user-defined scanning path. At each scan location, the $Z$-axis moves to the prescribed position and holds the model stationary relative to the measurement plane before image acquisition begins. The number of image pairs acquired at each plane, the hold time before acquisition, and the full sequence of $Z$ positions are specified in MATLAB. After acquisition at one plane is completed, the $Z$-axis increments to the next prescribed position and the procedure is repeated until all measurement planes are collected. To avoid the influence of the driveshaft wake, the lower measurement planes are typically reconstructed by mirroring the corresponding upper-plane data across the horizontal midline. This automated procedure enables repeatable plane spacing and allows long scanning sequences to be performed without manual intervention. The $Z$-positioning accuracy is evaluated using the same commanded--measured position tracking procedure described in the system performance characterization.

The synchronization architecture is summarized in Fig.~\ref{fig:12}(b). The control PC specifies the water-tunnel flow speed, the commanded $Z$-axis position, and the experimental kinematic parameters, including motion amplitude and frequency when phase-locked measurements are required. Once the model reaches the target scan location, the control PC sends a start command to the Teensy controller through USB serial communication. In response, the Teensy generates a single TTL pulse with a width of 1~ms and sends it to the external-trigger input of the LaVision programmable timing unit (PTU). The rising edge of this pulse starts the acquisition sequence previously configured in the LaVision system. The PTU then controls the synchronized camera timing. Thus, the Teensy provides only the acquisition-start trigger; the camera frame rate, inter-frame timing, and number of acquired images are controlled by the LaVision acquisition system. The acquired images are returned to the control PC for PIV processing, while the CPS controls the scan position and experimental kinematics.

The reconstruction procedure is shown in Fig.~\ref{fig:12}(c). Which used a prescribed-pitch configuration, with the $\Theta$-axis operating in active mode to impose $\theta(t)=\theta_0\sin(2\pi f t)$ at $\theta_0=15^\circ$. The $Z$-axis also operated in active mode and served as the automated scanning traverse, while the remaining translational axes were held fixed. The flow speed and pitching frequency were selected to give $St=0.60$. Stereo-PIV measurements at successive $Z$ locations are processed as individual two-dimensional, three-component velocity planes. For periodic motions, acquisition is phase locked so that each plane corresponds to the same phase. The planes are then placed at their prescribed $Z$ locations to reconstruct the three-dimensional wake, from which vorticity and other derived quantities can be evaluated.

The reconstruction places scan position, structural kinematics, hydrodynamic loads, camera triggering, and phase-aligned three-component velocity fields on the same trial definition. Because the optical arrangement remains fixed, each layer shares the same camera--laser calibration.

\section{Conclusion}\label{sec:conclusion}
We developed and validated a four-axis cyber-physical experimental framework for reconfigurable fluid--structure interaction studies in a water tunnel. The platform provides three translational degrees of freedom and one rotational degree of freedom, each independently assignable to prescribed motion, load-responsive virtual dynamics, or a fixed constraint. A common position-command architecture integrates real-time force/torque acquisition, virtual-dynamic computation, Stateflow-based mode supervision, and automated experiment execution. The two principal control pathways were evaluated separately. Prescribed pitching measurements reproduced established thrust and power scaling trends over the tested conditions, while the active-heave/passive-pitch benchmark reproduced the expected frequency-dependent resonant response. Together, these results establish the prescribed-motion and force-driven virtual-dynamic pathways over the operating ranges examined in this study.

The same architecture was then reconfigured for substantially different experimental roles without redesigning the underlying motion and sensing system. Hybrid pitching demonstrated intra-cycle reassignment between prescribed and load-responsive dynamics; the vertical-axis turbine surrogate demonstrated coordinated multi-axis orbital and rotational motion; the locomotion configuration transferred the force-driven response to the streamwise direction while retaining prescribed lateral actuation; and the scanning stereoscopic PIV configuration used the vertical axis as an automated diagnostic traverse. These cases are not intended as complete optimization studies of the corresponding physical systems. Rather, they demonstrate that different FSI boundary conditions, axis couplings, and measurement requirements can be expressed within a common hardware and control framework by changing the roles and coordination of the available degrees of freedom. In this sense, the principal contribution is not the number of controlled axes alone, but the integration of independently assigned operating modes, force-driven dynamics, intra-trial reconfiguration, automated execution, and diagnostic positioning within a reusable experimental architecture.

The present implementation nevertheless defines clear operating limits. The accessible motion and virtual-dynamic ranges are bounded by actuator travel, speed, force or torque capacity, structural rigidity, and the inertia of the moving gantry and test article. Because the present formulation includes the physical moving mass in the effective inertia, it also imposes a lower bound on the virtual mass that can be represented using the current position-command approach. The platform emulates rigid-body degrees of freedom and does not directly reproduce distributed structural deformation. During passive-to-active switching, displacement blending reduces command discontinuities, but velocity and acceleration cannot generally be matched simultaneously, so finite transients may remain. Scanning stereoscopic PIV is sequential and therefore provides a quasi-three-dimensional reconstruction rather than an instantaneous volumetric measurement; its usable scan range is additionally constrained by the tunnel depth and proximity of the model to facility boundaries. Within these operating constraints, the validated architecture provides a reusable experimental basis for configurable FSI studies. Future application-specific investigations can therefore reference the present work for the underlying hardware, sensing, motion-control, virtual-dynamic, mode-switching, and data-acquisition methodology.

\printcredits

\bibliographystyle{cas-model2-names}
\bibliography{ref}

\end{document}